\documentclass[11pt]{article}
\usepackage{bbold}
\usepackage[centertags]{amsmath}
\usepackage[square, comma, sort&compress,numbers]{natbib}
\usepackage{array,multirow}
\numberwithin{equation}{section}
\usepackage{amssymb}
\usepackage{comment}
\usepackage{graphicx}
\usepackage{float}
\usepackage{color}
\usepackage{epsfig}
\usepackage{tikz,pgf}
\usetikzlibrary{shapes}
\usetikzlibrary{calc}
\usetikzlibrary{decorations.pathmorphing}
\usetikzlibrary{decorations.pathreplacing,shapes.misc}
\usetikzlibrary{positioning}
\usetikzlibrary{arrows}
\usetikzlibrary{decorations.markings}
\tikzset{snake it/.style={decorate, decoration=snake}}

\def\be{\begin{equation}}
\def\ee{\end{equation}}
\def\ba{\begin{array}}
\def\ea{\end{array}}
\def\te{\textstyle}
\newcommand{\bref}[1]{\textbf{\ref{#1}}}
\def\dps{\displaystyle}
\newcommand{\half}{\frac{1}{2}}
\def\tr{{\rm Tr}}

\def\la{\langle}
\def\ra{\rangle}

\def\dps{\displaystyle}

\def\cN{\mathcal{N}}

\usepackage{jheppub}
\makeatletter
\def\@fpheader{\vspace{-.1cm}}
\makeatother

\title{c-Recursion for multi-point superconformal blocks.  \\NS sector}

\author[a,b,c\,\dagger]{Vladimir\ Belavin,}
\note[$\dagger$]{Weston Visiting Professorship at Weizmann Institute.}
\author[d,e]{\;Roman\ Geiko}

\affiliation[a]{I.E. Tamm Department of Theoretical Physics, P.N. Lebedev Physical
Institute,\\ Leninsky ave. 53, 119991 Moscow, Russia}
\affiliation[b]{Department of Quantum Physics,
Institute for Information Transmission Problems, \\
Bolshoy Karetny per. 19, 127994 Moscow, Russia}
\affiliation[c]{Department of Particle Physics and Astrophysics, Weizmann Institute of Science,\\
Rehovot 7610001, Israel}
\affiliation[d]{National Research University Higher School of Economics, \\
Usacheva str. 6, 119048 Moscow, Russia}
\affiliation[e]{ Center for Advanced Studies,
Skolkovo Institute of Science and Technology,\\
143026 Moscow, Russia}
\emailAdd{belavin@lpi.ru}
\emailAdd{romangeiko@mail.ru}

\abstract{
We develop a recursive approach to computing Neveu-Schwarz conformal blocks associated with n-punctured Riemann surfaces.
This work generalizes  the results of \cite{Cho:2017oxl} obtained recently for the Virasoro algebra.  The method is based on the
analysis of the analytic properties of the superconformal blocks considered as functions of the central charge $c$. It consists of  two
main ingredients: the study of the singular behavior of the conformal blocks and the analysis of their asymptotic properties when $c$
tends to infinity. The proposed construction is applicable for computing multi-point blocks in different topologies. We consider some
examples for genus zero and one with different numbers of punctures. As a by-product, we propose a new way to solve the recursion
relations, which gives more efficient computational procedure and can  be applied to SCFT case as well as to pure Virasoro blocks.
}

\begin{document}

\maketitle
\flushbottom

\section{Introduction}
It is well known that conformal blocks  \cite{Belavin:1984vu} are needed for computing correlation functions in any CFT model. Given
that the space of local fields in the model consists of a set of irreducible representations of a symmetry algebra
$\mathcal{A}_\text{sym}$($\supset $ Virasoro algebra),  conformal blocks  are defined, using the concept of operator product expansion
(OPE), as the holomorphic contribution to the correlation function coming from particular sets of representations in the intermediate
OPE
channels. Among conformal blocks the four-point blocks play most prominent role because they are needed for performing conformal
bootstrap program (for the recent review, see \cite{Poland:2018epd}).

Most straightforwardly, conformal blocks can be expanded as a sum over irreducible representations (each consisting of a primary
operator and its descendants) by inserting a complete set of states in each intermediate channel. For a given set of OPE channels this
expansion corresponds to some special way of gluing three-punctured spheres (the so-called  pant decomposition of the conformal blocks)
and gives the expression of the blocks in terms of the matrix elements of chiral vertex operators between two states (primary or
descendant). However this way has obvious technical restriction because it requires inversion of the matrix of scalar products of the
descendant states, which in general is not diagonal and grows rapidly with the descendants' level.\footnote{Using AGT correspondence
\cite{Alday:2009aq} the problem of non-diagonality can be solved for  (super-)Virasoro algebra by extending \cite{Belavin:2011pp} the
symmetry algebra and introducing the so-called integrable  basis  \cite{Alba:2010qc,Belavin:2011js}, however the rapid growth problem
still remains.}

Another approach to computing conformal blocks  is based on the study of their analytic properties. It  has been initially invented in
\cite{Zamolodchikov:1985ie,Zamolodchikov1987}  for computing four-point Virasoro blocks on the sphere by means of two types of
recursions: the so-called $c$-recursion ($c$ being the central charge parameter) and elliptic recursion (defined in terms of the
intermediate conformal dimension $h$), exploiting analytic properties of the conformal blocks, considered as functions of $c$ and  $h$
respectively. This  allowed to verify  the crossing relation for four-point correlation  functions and thus to perform the  bootstrap
program  in the Liouville theory \cite{Zamolodchikov:1995aa}. Later these results  have been extended  to the case of $\cN=1$
super-symmetric  Liouville theory \cite{Hadasz:2006qb,Belavin:2006zr,Belavin:2007gz}. The generalization to the torus  has been
considered in \cite{Hadasz:2009db,Hadasz:2012im}, however only one-point blocks have become available through the recursion
constructions.

Meanwhile, the ability to compute conformal blocks is of interest from the AdS$_{d+1}$/CFT$_d$ perspective. One of the basic questions
here is what AdS objects  correspond to the boundary CFT conformal blocks. Significant progress in clarifying this question has been
made
in recent years
\cite{Hartman:2013mia,Asplund:2014coa,Fitzpatrick:2014vua,Hijano:2015rla,Alkalaev:2015wia,Alkalaev:2015lca,Hijano:2015qja,Hijano:2015zsa,Bhatta:2016hpz,Alkalaev:2016rjl,Belavin:2017atm,Kraus:2017ezw,Alkalaev:2017bzx,Nishida:2018opl,Bhatta:2018gjb,Hikida:2018eih}
for $d=2$. It was shown that in the large $c$ regime the so-called heavy-light\footnote{The term referring to a certain behavior \cite{Fitzpatrick:2015zha} of the
conformal dimensions in the limit $c\to \infty$. For other possible regimes in this context, see, e.g., \cite{Kusuki:2018wcv,Kusuki:2018nms}.} conformal blocks have clear
geometric interpretation in terms of  geodesic Witten diagrams. We notice that  $c$-recursion fits naturally into AdS$_3/$CFT$_2$
context, because the large central charge limit, which is relevant for the semiclassical approximation in the dual gravity path
integral,
corresponds to the regular part of the conformal block in the $c$-recursion construction, as will be explained below. While the previous
studies were based often on  the direct matrix elements computation, more efficient methods of the recursion construction can serve as a
useful tool  for the further analysis of the correspondence. We notice that the sub-leading corrections in $c$ for higher multi-point
correlation functions, especially on the torus, are basically not available from both the CFT and AdS sides, also the question of
supersymmetric extenision of the  AdS$_3/$CFT$_2$ correspondence remains almost unstudied (see, \cite{Fitzpatrick:2014oza,Chen:2016cms,Aharony:2016dwx}).

Recently, it was shown \cite{Cho:2017oxl} that Zamolodchikov's analysis of the Virasoro CFT can be also applied for computing  higher
multi-point correlators on the sphere and torus and, in principle, on higher genius Riemann surfaces. In this paper we develop analogous
techniques for multi-point blocks in $\cN=1$ super-Virasoro (SVir) field theory. We focus on the Neveu-Schwarz (NS) sector of the theory
(for precise definition see Section \bref{section.SuperVirasoroModule}).

The analytic properties with respect to $c$ give rise to the splitting of SVir blocks into two parts: the regular part, corresponding to
the limit $c\to \infty$ (the so-called light asymptotic), and the singular part, coming from (in general) simple poles, located at
degenerate values  of the intermediate conformal dimensions, $d_{r,s}(c)$. On the sphere the regular part is governed  by $osp(1|2)$
subalgebra of SVir, so that the light SVir blocks on the sphere are reduced to the blocks of $osp(1|2)$ (the so-called global blocks).
The light blocks on the torus are a bit more involved objects: they can be factorized on the  global $osp(1|2)$ blocks on the torus and
NS vacuum characters.

Another approach to compute Virasoro blocks, different from the mentioned above, has been proposed in \cite{Perlmutter:2015iya}.
This approach is based on decomposing contributions of conformal families into sums over modules growing from quasiprimary states. This
allows to express conformal blocks as a sum of light blocks with shifted intermediate dimensions. This representation  gives in fact
the
solution of $c$-recursion relation for the Virasoro 4-point block and provides in principle more efficient way to compute general
blocks.

In the present work we consider the $N$-point SVir conformal blocks on the sphere in the linear channel (including 4-point case) and
provide $c$-recursion formulas for them. Then, we consider non-linear OPE channel and  obtain $c$-recursion formulas for the
correspondind SVir block. Further, we provide the c-recursion for the n-point SVir block in a necklace channel on the torus (including
1-point case). Using the approach of \cite{Perlmutter:2015iya} we obtain the solutions of $c$-recursion relations  in the considered
cases. In Section~\bref{section.SuperVirasoroModule} we recall some facts about NS sector of $\cN=1$
super-Virasoro algebra and its modules.

In Sections~\ref{section.ConfBlocksSphere} and~\bref{section.ConfBlocksTorus} we analyze  the
recursion relations for superconformal blocks on the sphere and torus respectively. In Section~\bref{section.SolRecurRel} we discuss an
improvement, that allows to resolve $c$-recursion formulas. Some technical details are collected in the appendices.

\section{Preliminaries}
\label{section.SuperVirasoroModule}
In this section we briefly remind some facts about  NS sector of $\cN=1$ CFT, which are relevant for our purposes. For more systematic
exposition and the general landscape, see, e.g., \cite{Nakayama:2004vk,Belavin:2007gz}.

\subsection{NS sector fields}
The fields of $\cN=1$ CFT  belong to highest weight representations (modules)  of the superconformal
algebra\footnote{Because the object of our study is conformal blocks, from now on we neglect the second (antiholomorphic) copy of the
symmetry algebra and suppress the dependence on $\bar z$.}
\begin{eqnarray}
\label{NS}
\nonumber
\left[L_n,L_m\right] & = & (n-m)L_{m+n} +\frac{c}{8}n\left(n^2-1\right)\delta_{m+n,0}\;,
\\
\left[L_n,G_k\right] & = &\left(\frac{n}{2}-k\right)G_{n+k}\;,
\\
\nonumber
\left\{G_k,G_l\right\} & = & 2L_{k+l} + \frac{c}{2}\left(k^2 -\frac14\right)\delta_{k+l,0}\;,
\end{eqnarray}
where in the NS sector
\begin{equation*}
\begin{alignedat}{2}
&n,m\:\in Z\quad\text{and} \quad k,l\:\in\mathbb{Z}+\te\frac{1}{2}\;
\end{alignedat}
\end{equation*}
and $c$ is the central charge.

The highest weight vector or the primary state $|d\ra$ is defined as
\be
\begin{aligned}
L_{0}|d\ra&=d\,|d\ra\;,\\
L_{n}|d\ra&=0\;,~ n>0\;,\\
G_{k}|d\ra&=0\;,~ k>0\;,
\end{aligned}
\ee
where $d$ is the conformal dimension parameter.  In the Liouville-like $(\lambda,b)$-parametrization, the conformal weight of the
primary
state is given by
\be
d(\lambda)=\frac{(b+b^{-1})^2}{8}-\frac{\lambda^2}{8}\;,
\ee
and
\be\label{c-b}
c(b)=1+2(b+b^{-1})^2\;.
\ee

The module $\mathcal{H}_d$ is spanned by the states
\be
\mathcal L_{\vec k}\,|d\rangle=
G_{-l_1}...G_{-l_p}L_{-n_1}... L_{-n_m}|d \rangle\;,
\label{descendent}
\ee
where $0<l_1<...< l_p$ and $0<n_1\le ...\le n_m$. The grading with respect to $L_{0}$:
\be
L_0\, \mathcal L_{\vec k}\,|d\rangle=k\, \mathcal L_{\vec k}\,|d\rangle\;,\quad \text{with}~k=\sum_{i=1}^p l_i+\sum_{j=1}^{m}
n_j\;,
\ee
defines the level $k\in \mathbb{N}_0+\frac{1}{2}$ of the descendant state $\mathcal L_{\vec k}\,|d\rangle$. The states $|d\rangle$ and
$G_{-\half}|d\rangle$ correspond, respectively, to the (lower and upper) components of the primary supermultiplet $V_{d}(z)$ and
$\widetilde V_{d}(z)$. The conjugation is defined  by
\be\label{conj}
\langle d|\,\mathcal L_{\vec k}^{\dagger}=
\langle d |\,L_{n_m}... \,L_{n_1}G_{l_p}...\,G_{l_1}\;.
\ee
We denote by $B_{\vec k_1,\vec k_2}(d,c)$\footnote{We shall omit one or both arguments of $B_{\vec k_1,\vec k_2}(d,c)$, when it does not
lead to confusion.} the matrix of scalar products
\be
B_{\vec k_1,\vec k_2}=\la d| \mathcal L_{\vec k_1}^{\dagger}\mathcal L_{\vec k_2}|d\ra\;,
\ee
and fix the normalization condition $\la d|d\ra=1$.
The matrix of scalar products has block diagonal structure $B_{\vec k_1,\vec k_2}\sim \delta_{k_1,k_2}$ with the size $P_{NS}(k)$ of
$k$-th level block defined by the NS character
\be
\sum_{k\in \half \mathbb{N}_0}P_{NS}(k)q^{k}=\prod_{p=1}^\infty \frac{1+q^{p-\half}}{1-q^p}\;.
\ee

For the special values $d_{r,s}:=d(\lambda_{r,s})$, where $\lambda_{r,s}=rb+sb^{-1}$,  $r,s\in \mathbb{Z}_+$ and $r+s$ is even, the NS module
$\mathcal{H}_d$ becomes degenerated -- the $\frac{rs}{2}$-th level contains a singular vector $\chi_{r,s}$.\footnote{In what follows we
use the normalization of the singular vectors with coefficient 1 in front of $G_{-\half}^{rs} \nu_{r,s}$.}
Restricted on the $n$-th level, the Kac determinant is given by
\be
\det B^{(n)}=\prod_{1\le rs \le 2n}(d-d_{r,s})^{P_{NS}(n-\small\frac{rs}{2})}\;.
\ee
Hence, for $n\geq\frac{rs}{2}$,  in the module $\mathcal{H}_d$ with $d=d_{r,s}(c)$, the Kac determinant $\det B^{(n)}=0$.

\subsection{Matrix elements}
Below we denote  by $\nu_i$ and $V_i(z)\equiv V_{d_i}(z)$ the primary state $|d_i\rangle$ and the corresponding vertex operator. The
matrix elements of a general vertex operator can be regarded as three-linear forms:
\begin{align}
\rho(.~,~.~,~.):\mathcal{H}^{3}\to\mathbb{ C}\;, \quad
\rho(\mathcal{L}_{\vec k_1}\nu_1,\mathcal{L}_{\vec k_2}\nu_2,\mathcal{L}_{\vec k_3}\nu_3)=\la d_1|\mathcal{L}_{\vec k_1}^{\dagger}
(\mathcal{L}_{\vec k_2 }V_2(z))\mathcal{L}_{\vec k_3}|d_3\ra \Big|_{z\to 1}\;.
\end{align}
Here we assume the following normalization condition:
\be
\la d_1|V_2(z)|d_3\ra=z^{d_1-d_2-d_3}\;,\quad \la d_1|\widetilde{V}_2(z)|d_3\ra=z^{d_1-d_2-d_3-\half}\;.
\ee
The commutation relations between super-Virasoro generators and vertex operators:
\begin{align}
\label{NS}
&[L_n,V_{d}(z)]  =  z^n(z\partial_{z}+(n+1)d)V_{d}(z)\;,
\\
&[L_n,\tilde{V}_d(z)]  =  z^n\left(z\partial_{z}+(n+1)\left(d+\textstyle{\half}\right)\right)V_{d}(z)\;,
\\
&[G_k, V_{d}(z)] =  z^{k+\textstyle{\half}}\tilde V_{d}(z)\;,
\\
&\!\{G_k, \tilde V_{d}(z)\}  =  z^{k-\textstyle{\half}}(z\partial_{z}+d(2k+1)) V_{d}(z)\;,
\end{align}
allows one to express the matrix elements involving arbitrary descendants in terms of the two independent matrix elements
$\rho^{\alpha_2}(\nu_1,\nu_2,\nu_3)$, where $\alpha_2\in\{0,1\}$, defined as follows
\be
\rho^{0}(\nu_1,\nu_2,\nu_3):=\rho(\nu_1,\nu_2,\nu_3)\;\quad\text{and}\quad\rho^{1}(\nu_1,\nu_2,\nu_3):=\rho(\nu_1,G_{-\half}\nu_2,\nu_3)\;.
\ee

The matrix elements of singular vectors $\rho(\chi_{r,s}(c),~.~,~.)$ must vanish if superconformal  fusion rules $\nu_{r,s}\times
\nu_2\to \nu_3$  are satisfied and therefore can be expressed in terms of the so-called fusion polynomials
\cite{Hadasz:2006qb,Belavin:2006zr}
\be \label{Peven}
P_{r,s}^0(d_i,d_j;c)=\!\!\prod_{\substack{p=1-r\\\text{step 2}}}^{r-1}\prod_{\substack{q=1-s\\\text{step
2}}}^{s-1}\bigg(\frac{\lambda_i-\lambda_j+pb+qb^{-1}}{2\sqrt 2}\bigg)\bigg(\frac{\lambda_i+\lambda_j+pb+qb^{-1}}{2\sqrt 2}\bigg)\,,\,
p+q-(r+s)\in
4 \mathbb{Z}+2\;,
\ee
\be \label{Podd}
P_{r,s}^1(d_i,d_j;c)=\!\!\prod_{\substack{p=1-r\\\text{step 2}}}^{r-1}\prod_{\substack{q=1-s\\\text{step
2}}}^{s-1}\bigg(\frac{\lambda_i-\lambda_j+pb+qb^{-1}}{2\sqrt 2}\bigg)\bigg(\frac{\lambda_i+\lambda_j+pb+qb^{-1}}{2\sqrt 2}\bigg)\,,\,
p+q-(r+s)\in
4 \mathbb{Z}\;,
\ee
where subscript $c$ indicates the dependence on the central charge  parameter, which comes through the parameter $b$ according to
\eqref{c-b}. We denote\footnote{Other types of matrix elements with singular vectors, like $\rho (
\chi_{r,s}(c),\nu_i,G_{-\half}\nu_j)$, etc., are not linear independent and will not be relevant.}
\be
\rho^{\alpha} ( \chi_{r,s}(c),\nu_i,\nu_j)=\sigma_{r,s}^{\alpha}(d_i,d_j;c)\;,
\ee
then
\be
\sigma_{r,s}^{\alpha}(d_i,d_j;c)=\begin{cases}
P_{r,s}^{\alpha}(d_i,d_j;c)\;,\quad rs\in 2\mathbb{N}\;,\\
P_{r,s}^{t(\alpha)}(d_i,d_j;c)\;,\quad\,rs\in 2\mathbb{N}+1\;,
\end{cases}
\ee
where $t(0)=1$ and $ t(1)=0$. Taking into account reflection symmetry:
\be
\begin{aligned}
\label{reflection}
\rho(\mathcal{L}_{\vec k} \nu_1,\nu_2,\nu_3)&=\rho(\nu_3,\nu_2,\mathcal{L}_{\vec k} \nu_1)\;,\quad k\in 2\mathbb{N}_0\;,\\
\rho(\mathcal{L}_{\vec k} \nu_1,G_{-\half}\nu_2,\nu_3)&=\rho(\nu_3,G_{-\half}\nu_2,\mathcal{L}_{\vec k} \nu_1)\;,\quad k\in
2\mathbb{N}_0\;,\\
\rho(\mathcal{L}_{\vec k} \nu_1,\nu_2,\nu_3)&=\rho(\nu_3,\nu_2,\mathcal{L}_{\vec k} \nu_1)\;,\quad k\in
2\mathbb{N}_0+1\;,\\
\rho(\mathcal{L}_{\vec k} \nu_1,G_{-\half}\nu_2,\nu_3)&=-\rho(\nu_3,G_{-\half}\nu_2,\mathcal{L}_{\vec k} \nu_1)\;,\quad k\in
2\mathbb{N}_0+1\;,
\end{aligned}
\ee
which follows from the conjugation relation \eqref{conj} and the minus sign in the last relation is due to anticommutativity of the odd
operators, one gets
\be
\rho ^{\alpha}( \nu_i,\nu_j,\chi_{r,s}(c))=S^{\alpha}_{rs}\sigma_{r,s}^{\alpha}(d_j,d_i;c)\;,
\ee
 where
\be
S^{0}_{n}=1\;,\quad S^{1}_n=(-1)^n\;.
\ee
Now we find the matrix element with two identical singular vectors $\mu^{\alpha}_{r,s} (d;c):=\rho^{\alpha} (
\chi_{r,s},\nu_d,\chi_{r,s})$,
which reads
\be
\mu^{\alpha} _{r,s}(d,c)=\begin{cases}
P^{\alpha}_{r,s}(d,  d_{r,s}+\te\frac{rs}{2};c)P_{r,s}^{\alpha}(d,d_{r,s}; c)\;,\quad
rs\in 2
\mathbb{N}\;,\\
 S_{rs}^{\alpha}\, P_{r,s}^{t(\alpha)}(d,  d_{r,s}+\frac{rs}{2}; c)P_{r,s}^{\alpha}(d,  d_{r,s}; c) \;,\quad
 rs\in 2 \mathbb{N}+1\;.
\end{cases}
\ee
Here and below we use the following factorization property
\be
\label{factorization}
\rho^{\alpha} (\mathcal{L}_{\vec k} \chi_{r,s},\nu_i,\nu_j)=\rho^{\alpha} (\mathcal{L}_{\vec k}
\nu_{d_{r,s}+\frac{rs}{2}},\nu_i,\nu_j)\rho^{\alpha} (\chi_{r,s},\nu_i,\nu_j)\;,
\ee
and, in particular, we use
\be
\label{factorization2}
\rho^{\alpha} (\mathcal{L}_{\vec k} \chi_{r,s},\nu_d, \mathcal{L}_{\vec k} \chi_{r,s})=\mu_{r,s}^{\alpha} (d;c)\,\rho^{\alpha}
(\mathcal{L}_{\vec
k} \nu_{d_{r,s}+\frac{rs}{2}},\nu_d,\mathcal{L}_{\vec k} \nu_{d_{r,s}+\frac{rs}{2}})\;.
\ee
\paragraph{$h$-parametrization.}
Alternatively, we can consider the matrix elements  of singular vectors $\rho(\chi_{r,s}(c),~.~,~.)$ as functions of
the conformal dimension $h$ of the primary state $\nu_h$  possessing  the singular descendant. To this end we use
\be
c_{r,s}(h)=1+2(b_{r,s}+b_{r,s}^{-1})^2\;,
\ee
and
\be
h=\frac{(b_{r,s}+b_{r,s}^{-1})^2}{8}-\frac{(rb_{r,s}+sb_{r,s}^{-1})^2}{8}\;.
\ee
We define $P_{r,s}^{\alpha}(d_i, d_j; h):=P_{r,s}^{\alpha}(d_i, d_j; c_{r,s}(h))$  by means of \eqref{Peven} and
\eqref{Podd} with $b$-parameter replaced by $b_{r,s}(h)$. The matrix elements in the $h$-parametrization are given by
\be
\sigma_{r,s}^{\alpha}(d_i,d_j;h):=\sigma_{r,s}^{\alpha}(d_i,d_j;c_{r,s}(h))\;, \qquad \mu _{r,s}^{\alpha}(d;h):=\mu_{r,s}^{\alpha}
(d;c_{r,s}(h)) \;.
\ee

\section{Blocks on the sphere}
\label{section.ConfBlocksSphere}
In this and the subsequent sections we use the following convention: the matrix elements contributing to the conformal block correspond
to the vertices of the associated dual diagram,
the vertex operators correspond to vertical lines, while asymptotic
states  correspond to horizontal lines. We denote by $\alpha_i=0,1$ -- lower and upper components of $\nu_i$
respectively, where $i$ numerates external lines. To be more compact, we suppress dependence on $c$ and conformal
dimensions, as well as coordinate dependence, of the conformal blocks. Also we suppress obvious conformal prefactors. We will use unique notation $F$, $G$ for blocks considered in
each subsection, referring to the corresponding diagrammatic representation.

\subsection{Previous results: 4-point blocks}
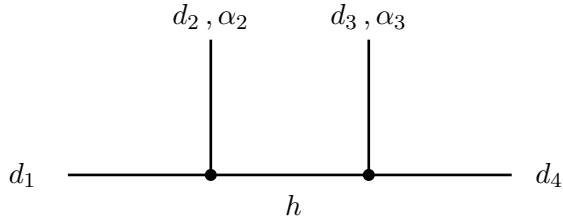
\begin{figure}[H]
\begin{center}
\begin{tikzpicture}

\draw [line width=1pt] (31.5,0) -- (37.4,0);
\draw [line width=1pt] (33.4,0) -- (33.4,1.8);
\draw [line width=1pt] (35.5,0) -- (35.5,1.8);


\draw (30.9,-0) node {$ d_1$};
\draw (33.4,2.1) node {$ d_{2}\,,\alpha_2$};
\draw (35.5,2.1) node {$ d_{3}\,,\alpha_3$};
\draw (37.9,0) node  {$ d_{4}$};

\draw (34.5,-0.4) node {$h$};


\fill  (33.4,0) circle (0.8mm);
\fill  (35.5,0) circle (0.8mm);

\end{tikzpicture}
\end{center}
\caption{Dual diagram for basic $4$-point superconformal blocks on the sphere.}
\label{Fig:4block}
\end{figure}

First we recapitulate the $c$-recursive representation for 4-point superconformal blocks, which has been developed in
\cite{Hadasz:2006qb,Belavin:2006zr}. Using super-projective invariance, we can chose among 16 blocks of lower and upper components of
primary supermultiplets (in the given OPE channel)
 four linear independent blocks according to the diagram depicted in Fig.~\bref{Fig:4block}.
\noindent These blocks have the following series expansion
\be
F=\sum_{n\in\half \mathbb{N}_0} F_n z^n\;,
\ee
where $z$ is anharmonic ratio of the holomorhic coordinates (in our convention $\nu_1,\nu_2,\nu_4$ stand at $\infty,1,0$ respectively)
and the coefficients are given by
\be
\label{4pt}
F_{n}=\sum_{\substack{k= m,\\m=n}}B^{\vec k,\vec m}\rho^{\alpha_2}(\nu_1,\nu_2,\mathcal{L}_{\vec k}\nu_{h})\rho^{\alpha_3}(\mathcal L
_{\vec
m}\nu_{h},\nu_3,\nu_{4})\;.
\ee
Here the sum goes over (half-)integer partitions and $B^{\vec k,\vec m}$ is the inverse Gram matrix.

The determinant of the Gram matrix has zeros at $h=d_{r,s}$. We consider the residues of the conformal block at the corresponding poles.
The coefficients $F_n$ for $n\ge\frac{rs}{2}$ have poles at $h=d_{r,s}$.
In the limit $h \to d_{r,s}$ the matrix elements entering \eqref{4pt} are non-singular and factorize according to \eqref{factorization}.
The singular vector $\chi_{r,s}$ is obtained by applying $\mathcal{L}_{r,s}$ to the degenerated state $\nu_{r,s}$. We denote by
$\chi^{h}_{r,s}$ the state, which is obtained by applying $\mathcal{L}_{r,s}$ to $\nu_h$ and which is the only one giving contribution
to
the residue of $B^{\vec k,\vec m}$. Using factorization property \eqref{factorization}, we get
\begin{multline}
\lim_{h\to d_{r,s}}(h-d_{r,s})\la \chi^{h}_{r,s}|\mathcal{L}_{\vec k}^{\dagger}\mathcal{L}_{\vec m}|\chi^{h}_{r,s}\ra^{-1} \\= \la
\nu_{d_{r,s}+\te\frac{rs}{2}}|\mathcal{L}_{\vec k}^{\dagger}\mathcal{L}_{\vec m}|\nu_{d_{r,s}+\te\frac{rs}{2}}\ra^{-1} \lim_{h\to
d_{r,s}}(h-d_{r,s})\la \chi^{h}_{r,s}|\chi^{h}_{r,s}\ra^{-1}=B^{\vec k,\vec m}(d_{r,s}+\te\frac{rs}{2})A_{r,s}(c)\;,
\end{multline}
where the coefficient
\be
A_{r,s}(c)=\half\prod_{p=1-r}^r\prod_{q=1-s}^{s}\bigg(\frac{1}{\sqrt{2}}\bigg(pb+qb^{-1}\bigg)\bigg)^{-1},\quad p+q\in 2\mathbb{Z},\quad
(p,q)\ne(0,0),(r,s)\;.
\ee
Assembling all the ingredients we get the following relation
\be
\lim_{h\to d_{r,s}}(h-d_{r,s})F=z^{\frac{rs}{2}}A_{r,s}(c)\,S_{rs}^{\alpha_2}\,\sigma_{r,s}^{\alpha_2}
(d_2,d_1;c)\,\sigma_{r,s}^{\alpha_3}
(d_3,d_4;c)\,F(h\rightarrow d_{r,s}+\te \frac{rs}{2},c)\;.
\ee
Considering the residues of the conformal block at $c=c_{r,s}(h)$, we have
\begin{multline}
\lim_{c\to c_{r,s}}(c-c_{r,s}(h))F=z^{\frac{rs}{2}}J_{r,s}(h)A_{r,s}(h)\,S_{rs}^{\alpha_2}\,\sigma_{r,s}^{\alpha_2}
(d_2,d_1;h)\,\sigma_{r,s}^{\alpha_3}
(d_3,d_4;h)\\
\times F(h\rightarrow h+\te \frac{rs}{2},c\rightarrow c_{r,s}(h))\;,
\end{multline}
where Jacobian  $J_{r,s}:=-\frac{\partial c_{r,s}}{\partial h}$ and
\be
A_{r,s}(h)=\half\prod_{p=1-r}^r\prod_{q=1-s}^{s}\bigg(\frac{1}{\sqrt{2}}\bigg(pb_{r,s}+qb^{-1}_{r,s}\bigg)\bigg)^{-1},\quad p+q\in
2\mathbb{Z},\quad (p,q)\ne(0,0),(r,s)\;.
\ee

In order to obtain the regular part or the light block we take the large  $c$ limit of the conformal block keeping all the dimensions
fixed. The contribution of the global subalgebra is dominating in this limit and the light NS super-Virasoro block $G$ is nothing but
the
conformal block associated to the $osp(1|2)$ subalgebra
 \begin{align}
  \label{4ptglobal}
\nonumber G=\sum_{n\in \half\mathbb{N}_0}G_{n}z^{n}
&=\sum_{n\in \mathbb{N}_0}\sum_{\,\,\,\,\beta\in
\{0,1\}}\!\!\!\!\!\!\!\frac{\rho^{0,\alpha_2,\beta}(\nu_1,\nu_2,L_{-1}^n\nu_{h})\rho^{\beta,\alpha_3,0}(L_{-1}^n\nu_{h},\nu_3,\nu_4)}{
\rho^{\beta,0,\beta}(L_{-1}^n\nu_{h},\mathbb{1},L_{-1}^n\nu_{h})}\,z^{n+\small\frac{\beta}{2}}\\
&=\sum_{n\in \mathbb{N}_0}\sum_{\beta\in \{0,1\}}\frac{(h +\te\frac{\beta}{2} +d_2 -d_1)_n \,(h +\te\frac{\beta}{2}+ d_3 -d_4)_n}{
n!\,(2h)_{n+\beta} }\,  z^{n+\small\frac{\beta}{2}}\;,
\end{align}
where $\mathbb{1}$ stands for the unity operator with $d=0$. The definition of $osp(1|2)$ algebra and the explicit global matrix elements
can
be found\footnote{For another approach to the analysis of the light asymptotic, based on AGT correspondence,
see~\cite{Poghosyan:2017qdv}.} in Appendix~\bref{global}. For the four-point block's coefficients  we get the following recursion
\be
\label{4pt}
F_{n}=G_{n}+\sum_{\substack {r\ge2,s\ge1,\\r+s\in 2\mathbb{N}}}^{rs\le
2n}\frac{R_{r,s}(h)}{c-c_{r,s}(h)}F_{n-\textstyle{\frac{rs}{2}}}(h\rightarrow h+\te\frac{rs}{2},c\rightarrow c_{r,s}(h))\;,
\ee
where
\be
\label{res4pt}
R_{r,s}(h)=J_{r,s}(h)A_{r,s}(h)\,S_{rs}^{\alpha_3}\,\sigma_{r,s}^{\alpha_2} (d_1,d_2;h)\,\sigma_{r,s}^{\alpha_3} (d_4,d_3;h)\;.
\ee

\subsection{Linear channel blocks}
\label{lin-channel}
\begin{figure}[H]
\begin{center}
\begin{tikzpicture}

\draw [line width=1pt] (31.6,0) -- (39.1,0);
\draw [line width=1pt] (33.4,0) -- (33.4,1.8);
\draw [line width=1pt] (35.4,0) -- (35.4,1.8);
\draw [line width=1pt] (37.4,0) -- (37.4,1.8);


\draw (31.2,-0) node {$ d_1$};
\draw (33.4,2.1) node {$ d_{2}, \alpha_2$};
\draw (35.5,2.1) node {$ d_{3}, \alpha_3$};
\draw (37.4,2.1) node {$ d_{4}, \alpha_4$};
\draw (39.5,0) node  {$ d_{5}$};

\draw (34.5,-0.4) node {$h_1$};
\draw (36.5,-0.4) node {$h_2$};


\fill  (33.4,0) circle (0.8mm);
\fill  (35.4,0) circle (0.8mm);
\fill  (37.4,0) circle (0.8mm);

\end{tikzpicture}
\end{center}
\caption{$5$-point superconformal block in the linear channel.}
\label{Fig:5block}
\end{figure}
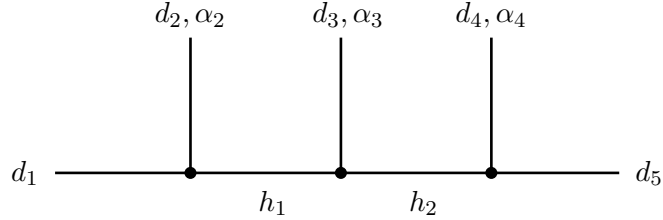

\paragraph{5-point block in the linear channel.}
Here we derive the recursion formulas for the 5-point conformal block in the linear OPE channel represented in Fig. \bref{Fig:5block}.
\noindent In this case  using  super-projective invariance we can fix $\nu_1$ and $\nu_5$ to be lower primary components and to put
operators $\nu_1, \nu_2,\nu_5$ at positions $\infty,1,0$ respectively. It is possible to construct a local parametrization of the moduli
space of Riemann sphere with $N$-punctures (see, e.g.,
\cite{Alkalaev:2015fbw,Cho:2017fzo}), so that conformal blocks have Taylor series expansion in these parameters. The conformal block
reads
\begin{multline}
F=\sum_{n_1,n_2\in \half N_0}F_{n_1,n_2}q_1^{n_1}q_2^{n_2}
=\sum_{\vec k_{1,2},\,\vec p_{1,2}}q_1^{k_1}q_2^{k_2}\,B^{\vec k_1,\vec p_1}(h_1)B^{\vec k_2,\vec p_2}(h_2)\times\\
\times\rho^{\alpha_2}(\nu_1,\nu_2,\mathcal{L}_{\vec k_1}\nu_{h_1})
\rho^{\alpha_3}(\mathcal{L}_{\vec p_1}\nu_{h_{2}},\nu_{3},\mathcal{L}_{\vec k_2}\nu_{h_{2}})
\rho^{\alpha_4}(\mathcal{L}_{\vec p_2}\nu_{h_{2}},\nu_{4},\nu_{5})\;,
\end{multline}
where, as explained in Appendix~\bref{Plumbing}, $q_1=z_3,\, q_2=\frac{z_4}{z_3}$. Analyzing singular behavior of the conformal block in the both intermediate channels, we get
\begin{multline}
F_{n_1,n_2}=G_{n_1,n_2}
+\sum_{\substack{r\ge2,s\ge1,\\r+s\in 2\mathbb{N}}}^{rs\le
2n_1}\frac{R^1_{r,s}(h_1)}{c-c_{r,s}(h_1)}F_{n_1-\te\frac{rs}{2},n_2}(h_1\rightarrow h_{r,s}+\textstyle{\frac{rs}{2}},c\rightarrow
c_{r,s})\\
+\sum_{\substack{r\ge2,s\ge1,\\r+s\in 2\mathbb{N}}}^{rs\le
2n_2}\frac{R^2_{r,s}(h_2)}{c-c_{r,s}(h_2)}F_{n_1,n_2-\te\frac{rs}{2}}(h_2\rightarrow h_{r,s}+\textstyle{\frac{rs}{2}},c\rightarrow
c_{r,s})\;,
\end{multline}
where coefficients $R^{i}_{r,s}$ are the following
\be
\begin{aligned}
R^1_{r,s}(h_1)=J_{r,s}(h_1)A_{r,s}(h_1)\,S_{rs}^{\alpha_2}\,\sigma_{r,s}^{\alpha_2} (d_2,d_1;h_1)\,\sigma_{r,s}^{\alpha_3}
(d_3,h_2;h_1)\;,\\
R^2_{r,s}(h_2)=J_{r,s}(h_2)A_{r,s}(h_2)\,S_{rs}^{\alpha_3}\,\sigma_{r,s}^{\alpha_3} (d_3,h_1;h_2)\,\sigma_{r,s}^{\alpha_4}
(d_4,d_5;h_2)\;.
\end{aligned}
\ee
Using the results of Appendix~\bref{global}, we find explicit coefficients of the 5-point light block
\begin{align}
\nonumber G&=\!\!\sum_{k,m\in  \half\mathbb{N}_0}G_{k,m}q_1^{k}q_2^{m}=
\sum_{k,m\in  \mathbb{N}_0}\sum_{\beta_{1,2}\in \{0,1\}} q_1^{k+\frac{\beta_1}{2}}\!\!q_2^{m+\frac{\beta_2}{2}} \times
\nonumber\\
&\hspace{2.3cm}
\times\frac{\rho^{0,\alpha_2,\beta_1}(\nu_1,\nu_2,L_{-1}^k\nu_{h_1})\rho^{\beta_1,\alpha_3,\beta_2}(L_{-1}^k\nu_{h_1},\nu_3,L_{-1}^m\nu_{h_2})\rho^{\beta_2,\alpha_4,0}(L_{-1}^m\nu_{h_2},\nu_4,\nu_5)}{
\rho^{\beta_1,0,\beta_1}(L_{-1}^k\nu_{h_1},\mathbb{1},L_{-1}^k\nu_{h_1})\,
\rho^{\beta_2,0,\beta_2}(L_{-1}^m\nu_{h_2},\mathbb{1},L_{-1}^m\nu_{h_2}) }
\nonumber\\
&=\!\sum_{k,m\in  \mathbb{N}_0}\sum_{\beta_{1,2}\in \{0,1\}}\!\!\!\!\!q_1^{k+\frac{\beta_1}{2}}q_2^{m+\frac{\beta_2}{2}}\frac{(h_1
+\te\frac{\beta_1}{2} +d_2 -d_1)_k \,(h_2 +\te\frac{\beta_2}{2}+ d_4 -d_5)_m}{ k!\,m!\,(2h_1)_{k+\beta_1} \,(2h_2)_{m+\beta_2}}\,
\tau_{k,m}^{\beta_1,\alpha_3,\beta_2}(h_1,d_3,h_2)\;.
\end{align}

\paragraph{$N$-point block in the linear channel.}
Below we generalize the previous consideration to $N$-point block in the linear channel.
\begin{figure}[H]
\begin{center}
\begin{tikzpicture}\label{Nblock}
\draw [line width=1pt] (30,0) -- (35,0);
\draw [line width=1pt] (38,0) -- (43,0);
\draw [line width=1pt] (32,0) -- (32,2);
\draw [line width=1pt] (34,0) -- (34,2);
\draw [line width=1pt] (39,0) -- (39,2);
\draw [line width=1pt] (41,0) -- (41,2);
\draw (29.5,-0) node {$ d_1$};
\draw (43.5,-0) node {$ d_N$};
\draw (32,2.3) node {$ d_{2}, \alpha_2$};
\draw (34,2.3) node {$ d_{3}, \alpha_3$};
\draw (41,2.3) node {$ d_{N-1}, \alpha_{N-1}$};
\draw (39,2.3) node {$ d_{N-2}, \alpha_{N-2}$};
\draw (33,-0.3) node {$h_1$};
\draw (40,-0.3) node {$h_{N-3}$};
\fill  (36.5,0) circle (0.3mm);
\fill  (36.7,0) circle (0.3mm);
\fill  (36.3,0) circle (0.3mm);

\fill  (32,0) circle (0.8mm);
\fill  (34,0) circle (0.8mm);
\fill  (39,0) circle (0.8mm);
\fill  (41,0) circle (0.8mm);
\end{tikzpicture}
\end{center}
\caption{$N$-point conformal block in the linear channel.}
\label{Fig:Nblock}
\end{figure}
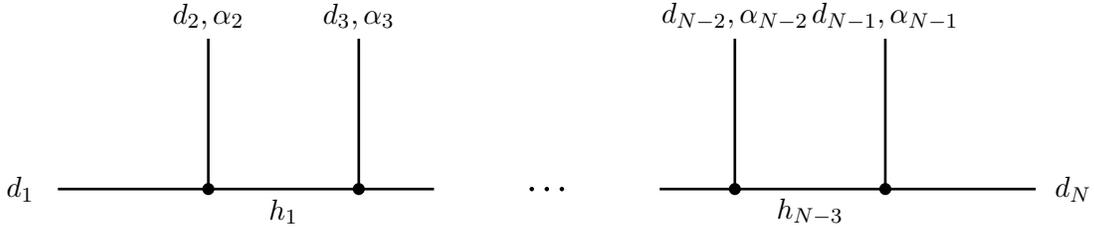
\noindent From Fig. \bref{Fig:Nblock} we read off the conformal block
\begin{multline}
F=\sum_{n_i\in \half N_0}F_{n_1,...,n_{N-3}}q_1^{n_1}...\,q_{N-3}^{n_{N-3}}
=\sum_{\vec k_i,\vec p_i}q_1^{k_1}...\,q_{N-3}^{k_{N-3}}\,B^{\vec k_1,\vec p_1}(h_1)...\,B^{\vec k_{N-3},\vec
p_{N-3}}(h_{N-3})\times\\
\times\rho^{\alpha_2}(\nu_1,\nu_2,\mathcal{L}_{\vec k_1}\nu_{h_1})...\,\rho^{\alpha_{N-1}}(\mathcal{L}_{\vec
p_{N-3}}\nu_{h_{N-3}},\nu_{N-1},\nu_{N})\;,
\end{multline}
where moduli parameters are related to the positions of the vertex operators through
\be
\label{Nmodules}
q_1=z_{3},\quad q_i=\frac{z_{i+2}}{z_{i+1}},\quad 2\le i\le N-3\;
\ee
and we fix $z_1=\infty, z_2=1, z_N=0$ (see Appendix~\bref{Plumbing}, Fig.~\bref{Fig:Nplumb}).

One gets the following recursion representation
\begin{multline}
F_{n_1,...,n_{N-3}}=G_{n_1,...,n_{N-3}}\\
+\sum_{i=1}^{N-3}\sum_{\substack{r\ge2,s\ge1,\\r_i+s_i\in 2\mathbb{N}}}^{r_is_i\le
2n_i}\frac{R^{i}_{r_i,s_i}(h_i)}{c-c_{r_i,s_i}(h_i)}
F_{n_1,...,n_i- \frac{r_i s_i}{2},...,n_{N-3}}(h_i\rightarrow h_{r_i,s_i}+\te\frac{r_i s_i}{2},c\rightarrow c_{r_i,s_i}(h_i))\;,
\end{multline}
were the coefficients
\begin{align}
&R^1_{r,s}(h_1)=J_{r,s}(h_1)A_{r,s}(h_1)\,S_{rs}^{\alpha_2}\,\sigma _{r,s}^{\alpha_2}(d_2,d_1;h_1)\,\sigma_{r,s}
^{\alpha_3}(d_3,h_2;h_1)\;, \\
&R^i_{r,s}(h_i)=J_{r,s}(h_i)A_{r,s}(h_i)\,S_{rs}^{\alpha_{i+2}}\,\sigma
_{r,s}^{\alpha_{i+2}}(d_{i+2},h_{i-1};h_i)\,\sigma_{r,s}^{\alpha_{i+3}}
(d_{i+3},h_{i+1};h_i)\;,~~1< i < N-3\;,\nonumber\\
&R^{N-3}_{r,s}(h_{N-3})=J_{r,s}(h_{N-3})A_{r,s}(h_{N-3})\,S_{rs}^{\alpha_{N-2}}\,\sigma_{r,s}^{\alpha_{N-2}}
(d_{N-2},h_{N-4};h_{N-3})\,\sigma_{r,s}^{\alpha_{N-1}}
(d_{N-1},d_N;h_{N-3})\nonumber
\end{align}
and $G_{n_1,...,n_{N-3}}$ is constructed from the matrix elements listed in Appendix~\bref{global}.

\subsection{Non-linear channel  blocks}
\label{nonlin-channel}
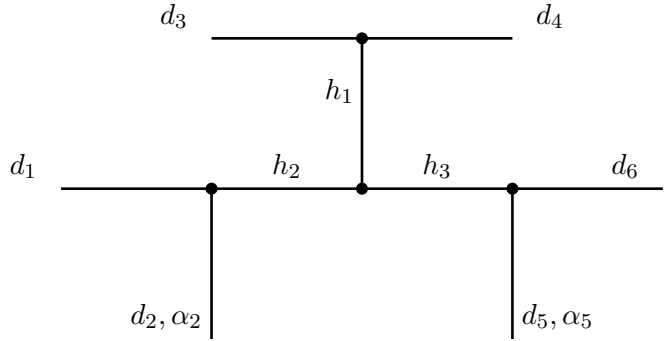
\begin{figure}[H]
\begin{center}
\begin{tikzpicture}
\draw [line width=1pt] (30,0) -- (34,0);
\draw [line width=1pt] (32,0) -- (32,-2);
\draw [line width=1pt] (28,-2) -- (36,-2);
\draw [line width=1pt] (30,-2) -- (30,-4);
\draw [line width=1pt] (34,-2) -- (34,-4);

\draw (29.5,0.3) node {$ d_3$};
\draw (34.5,0.3) node {$ d_4$};
\draw (27.5,-1.7) node {$ d_1$};
\draw (29.4,-3.7) node {$ d_2, \alpha_{2}$};
\draw (35.5,-1.7) node {$ d_6$};
\draw (34.6,-3.7) node {$ d_5, \alpha_{5}$};
\draw (31.7,-0.7) node {$ h_1$};
\draw (31.0,-1.7) node {$ h_2$};
\draw (33.0,-1.7) node {$ h_3$};
\fill  (32.0,0) circle (0.8mm);
\fill  (32.0,-2) circle (0.8mm);
\fill  (30.0,-2) circle (0.8mm);
\fill  (34.0,-2) circle (0.8mm);
\end{tikzpicture}
\end{center}
\caption{6-point conformal block in the non-linear channel.}
\label{Fig:6block}
\end{figure}

We fix the OPE channel according to the diagram in Fig.~\bref{Fig:6block}.
\noindent The  block reads
\begin{multline}
F=\sum_{\vec k_i,\vec p_i}
q_{1}^{p_1}q_2^{p_2}q_3^{p_3}\,B^{\vec k_1,\vec p_1}(h_1)B^{\vec k_2,\vec p_2}(h_2)B^{\vec k_3,\vec p_3}(h_3)\times\\
\times\rho(\nu_3,\mathcal{L}_{\vec
k_1}\nu_{h_1},\nu_4)
 \rho(\mathcal{L}_{\vec p_2}\nu_{h_2},\mathcal{L}_{\vec p_1}\nu_{h_1},\mathcal{L}_{\vec p_3}\nu_{h_3})
\rho^{\alpha_2}(\nu_{1},\nu_{2},\mathcal{L}_{\vec k_2}\nu_{h_2})
\rho^{\alpha_5}(\mathcal{L}_{\vec k_3}\nu_{h_3},\nu_{6},\nu_{5})\;,
\end{multline}
where the sum goes over the set of half-integer partitions. We fix operators $\nu_1, \nu_2, \nu_6$ to be located at $\infty,1,0$
respectively, and, as explained in Appendix~\bref{Plumbing} (see Fig.~\bref{Fig:plumb6pt}), the moduli of the punctured Riemann sphere are expressed in terms of the positions of the remaining operators as
follows:
$q_1=1-z_4,\;q_2=\frac{1}{z_3},\;q_3=z_5$.
Exploiting the pole decomposition in each intermediate  channel, we get the following recursion
\begin{multline}
F=G+\sum_{\substack{r\ge2,s\ge1\\r+s\in 2\mathbb{N}}}\frac{q_1^{\textstyle{\frac{r s}{2}}}R^1_{r,s}}{c-c_{r,s}(h_1)}F(h_1\rightarrow
h_1+\textstyle{\frac{r
s}{2}},c\rightarrow c_{r,s}(h_1))\\
+\sum_{\substack{r\ge2,s\ge1\\r+s\in 2\mathbb{N}}}\frac{q_2^{\textstyle{\frac{r s}{2}}}R^2_{r,s}}{c-c_{r,s}(h_2)}F(h_2\rightarrow
h_2+\textstyle{\frac{r
s}{2}},c\rightarrow c_{r,s}(h_2))\\
+\sum_{\substack{r\ge2,s\ge1\\r+s\in 2\mathbb{N}}}\frac{q_3^{\textstyle{\frac{r s}{2}}}R^3_{r,s}}{c-c_{r,s}(h_3)}F(h_3\rightarrow
h_3+\textstyle{\frac{r
s}{2}},c\rightarrow c_{r,s}(h_3))\;,
\end{multline}
where
\be
\begin{aligned}
R^1_{r,s}&=A_{r,s}(h_1)\,J_{r,s}(h_1)\,\sigma_{r,s}^{\alpha_4}(d_3,d_4;h_1)\,\sigma_{r,s}^{0}(h_2,h_3;h_1)\;,\\
R^2_{r,s}&=A_{r,s}(h_2)\,J_{r,s}(h_2)\,S_{r,s}^{\alpha_2}\,\sigma_{r,s}^{\alpha_2}(d_1,d_2;h_2)\,\sigma_{r,s}^{0}(h_3,h_1;h_2)\;,\\
R^3_{r,s}&=A_{r,s}(h_3)\,J_{r,s}(h_3)\,\sigma_{r,s}^{\alpha_5}(d_5,d_6;h_3)\,\sigma_{r,s}^{0}(h_1,h_2;h_3)\;.
\end{aligned}
\ee
One can compare these results to the AGT construction for the non-linear Virasoro block found
in~\cite{Hollands:2011zc}.

\section{Blocks on the torus}
\label{section.ConfBlocksTorus}
\subsection{Previous results: 1-point blocks}
\label{1pt}
\begin{figure}[H]
\begin{center}
\begin{tikzpicture}
\draw [line width=1pt] (30,0) -- (30,1.5);

\draw (29.5,1.3) node {$ d, \alpha$};
\draw (29.0,-2.1) node {$ h$};
\draw [line width=1pt] (30.0,-1.0) circle (1.0);
\fill  (30.0,0) circle (0.8mm);

\end{tikzpicture}
\end{center}
\caption{Torus $1$-point conformal block.}
\label{Fig:1torusblock}
\end{figure}
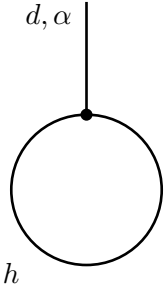

Here we apply the recursive  approach to computing NS 1-point conformal blocks on the torus.
The corresponding graph is represented in Fig.~\bref{Fig:1torusblock}.
\noindent By definition, the block is given by
\be
\begin{aligned}
\label{torus}
&F=\sum_{n\in \half \mathbb{N}_0}^{\infty}F_n\;,\\
&F_{n}=\!\!\sum_{k=n}B_{n}^{\vec k, \vec k}(h)\rho^{\alpha} (\mathcal{L}_{\vec k} \nu_h,\nu_d, \mathcal{L}_{\vec k} \nu_h)\;,
\end{aligned}
\ee
where $q=e^{2\pi i \tau}$ and $\tau$ is the modular parameter of the torus (see Appendix~\bref{Plumbing}, Fig.~\bref{Fig:plumb1pttor}).

The residue of the conformal block is given by
\be
\lim_{h\to d_{r,s}}(h-d_{r,s})F=q^{\frac{rs}{2}}A_{r,s}(c)\,\sum_{k\in \half \mathbb{N}_0} z^{k}B^{\vec k, \vec
k}(d_{r,s}+\textstyle{\frac{rs}{2}})\,\mu_{r,s}^{\alpha}(d;c)\rho^{\alpha}
(\mathcal{L}_{\vec k} \nu_{d_{r,s}+\frac{rs}{2}},\nu_d, \mathcal{L}_{\vec k} \nu_{d_{r,s}+\frac{rs}{2}})\;.
\ee
 After changing the variables one gets
 \begin{multline}
\lim_{c\to c_{r,s}(h)}(c-c_{r,s}(h))F=q^{\frac{rs}{2}}J_{r,s}(h)A_{r,s}(h)\,q^{\frac{rs}{2}}\\
\times \sum_{k\in \half \mathbb{N}_0} z^{k}B^{\vec k, \vec
k}(d_{r,s}+\textstyle{\frac{rs}{2}})\,\mu_{r,s}^{\alpha}(d;h)\rho^{\alpha}
(\mathcal{L}_{\vec k} \nu_{d_{r,s}+\frac{rs}{2}},\nu_d, \mathcal{L}_{\vec k} \nu_{d_{r,s}+\frac{rs}{2}})\;.
\end{multline}
 This defines the singular part of the pole decomposition. Now, to get $c$-recursion relations, it remains to find the light asymptotic
 of the block. In the full analogy with the Virasoro case \cite{Alkalaev:2016fok}, where the light 1-point block on the torus factorizes
 into the vacuum character and the global block, the NS light block splits into the NS vacuum character and the global  block of
 $osp(1|2)$ algebra~\cite{Alkalaev:2018qaz}
\be
\label{toruslight}
G=\prod_{n=2}^\infty\frac{1+q^{n-\half}}{1-q^n} \times F \big|_{osp(1|2)}=\sum_{n\in \half \mathbb{N}_0}^\infty G_{n}q^{n}\;.
\ee
The global block reads
\begin{multline}\label{torusglobal}
 F \big|_{osp(1|2)}=
\sum_{n\in\mathbb{N}_{0}}\sum_{\beta\in\{0,1\}}q^{n+\frac{\beta}{2}}\frac{\rho^{\beta,\alpha,\beta}(L_{-1}^{n}\nu_h,\nu_d,L_{-1}^{n}\nu_h)}{\rho^{\beta,0,\beta}(
L_{-1}^{n}\nu_h,\mathbb{1},L_{-1}^{n}\nu_h\ra}
=\sum_{n\in \mathbb{N}_0}\sum_{\beta\in\{0,1\}} q^{n+\frac{\beta}{2}}\frac{\tau^{\beta,\alpha,\beta}_{n,n}(h,d,h)}{n!(2h)_{n+\beta}}\;,
\end{multline}
where the explicit form of the matrix elements is given in Appendix~\bref{global}. Finally, we get the following recursion
relations for the one-point torus superblock coefficients
\be
\begin{aligned}
\label{1pttor}
&F_{n}=G_n+\sum_{\substack{r\ge2,s\ge1,\\r+s\in 2\mathbb{N}}}^{rs\le 2n}\frac{R_{r,s}(h)}{c-c_{r,s}(h)}F_{n- \frac{rs}{2}}(h\rightarrow
h+\te \frac{rs}{2},c\rightarrow c_{r,s})\;,\\
\end{aligned}
\ee
where the coefficients $G_n$ are defined in \eqref{toruslight}, \eqref{torusglobal} and the residue coefficients are
\be
\label{Rtor}
R_{r,s}(h)=J_{r,s}(h)A_{r,s}(h)\,\mu_{r,s}^{\alpha}(d;h)\;.\\
\ee
The lower coefficients of the block $F$ are collected in Appendix~\bref{1tor}.

\subsection{$N$-point blocks in the necklace channel}
\label{necklace-channel}
The generalization of the previous consideration on the $N$-point case is pretty straightforward. We define the block
\be
F=\sum_{\vec k_i,\vec p_i}q_1^{k_1}...\,q_N^{k_N}B^{\vec k_1,\vec p_1}(h_1)...\,B^{\vec k_N,\vec p_N}(h_N)\,\rho^{\alpha_1}(L_{\vec
p_1}\nu_1,\nu_{h_1},L_{\vec k_2}\nu_2)\,...\,\rho^{\alpha_N}(L_{\vec p_N}\nu_N,\nu_{h_N},L_{\vec k_1}\nu_1)\;,
\ee
corresponding to the diagram in Fig.~\bref{Fig:Ntorusblock} (for the definition of $q_i$, see Appendix \bref{Plumbing}, Fig.~\bref{Fig:Nplumbtorus}).
\begin{figure}[H]
\begin{center}
\begin{tikzpicture}
\draw [line width=1pt] (30,0) -- (30,2);
\draw [line width=1pt] (30.7,-0.3) --(32.1,1.1) ;
\draw [line width=1pt] (29.3,-0.3) -- (27.9,1.1);
\draw (29.5,1.7) node {$ d_1, \alpha_{1}$};
\draw (31.3,1) node {$ d_2, \alpha_{2}$};
\draw (30.5,0.2) node {$ h_1$};
\draw (29.5,0.2) node {$ h_N$};
\draw (27.6,0.7) node {$ d_N, \alpha_{N}$};
\draw [line width=1pt] (30.0,-1.0) circle (1.0);
\fill  (30.0,0) circle (0.8mm);
\fill  (30.0,-2) circle (0.8mm);
\fill  (31.0,-1) circle (0.8mm);
\fill  (29.0,-1) circle (0.8mm);
\fill  (30.7,-0.3) circle (0.8mm);
\fill  (30.7,-1.7) circle (0.8mm);
\fill  (29.3,-1.7) circle (0.8mm);
\fill  (29.3,-0.3) circle (0.8mm);

\fill  (32.0,-1) circle (0.4mm);
\fill  (31.98,-1.2) circle (0.4mm);
\fill  (31.98,-0.8) circle (0.4mm);

\end{tikzpicture}
\end{center}
\caption{Torus $N$-point conformal block in the necklace channel.}
\label{Fig:Ntorusblock}
\end{figure}
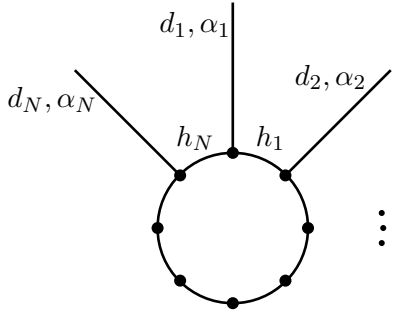
\noindent Using our general scheme, we get the following decomposition
\be
F=G+\sum_{i=1}^N  \sum_{\substack{r\ge2,s\ge1,\\r_i+s_i\in 2\mathbb{N}}} \frac{R^i_{r_i,s_i}}{c-c_{r_i,s_i}(h_i)} F(c\rightarrow
c_{r_i,s_i}(h_i),h_i\rightarrow h_i+\te\frac{r_i s_i}{2})\;,
\ee
where
\be
R^1=J_{r,s}(h_1)A_{r,s}(h_1)\,S_{rs}^{\alpha_1}\,\sigma_{r,s}^{\alpha_1}(d_1,h_{N};h_1) \,\sigma_{r,s}^{\alpha_2}(d_2,h_2;h_1)
\ee
and other $R$-coefficients are obtained by replacing  cyclically $h_{i}$, $d_i$ and $\alpha_i$. The light block is
\be
G=\prod_{n_{\small 1}=2}^{\infty}...\prod_{n_{\small N}=2}^{\infty}\frac{1+q_1^{n_{\small 1}-\half}...\,
q_N^{n_N-\half}}{1-q_1^{n_{\small
1}}...\,q_N^{n_{\small N}}}\times F \big|_{osp(1|2)}=\!\!\!\!\sum_{n_{i}\in \half \mathbb{N}_0}G_{n_{\small
1}...n_{\small
N}}q_1^{n_{\small 1}}...\,q_{\small N}^{n_{\small N}}\;,
\ee
with the global block constructed of the $osp(1|2)$ matrix elements (see Appendix \ref{global})
\be
\begin{aligned}
F \big|_{osp(1|2)}=\!\!\!\!\sum_{n_{i}\in \mathbb{N}_0}\sum_{\beta_i\in \{0,1\}}
\!\!\!\!\frac{\tau^{\beta_1,\alpha_2,\beta_1}(h_1,d_2,h_2)}{(2h_1)_{n_1+\beta_1}n_1!}\,...\,\frac{\tau^{\beta_N,\alpha_1,\beta_N}(h_N,d_1,h_1)}{(2h_N)_{n_{\small
N}+\beta_N}n_{\small N}!}q_1^{n_{\small
1}+\beta_1}...\,q_{\small N}^{n_{\small N}+\beta_N}\,.
\end{aligned}
\ee
\section{Solutions of the recursion relations}
\label{section.SolRecurRel}
\paragraph{Solution for 4-point blocks on the sphere.}
Here we describe an approach to the computation of the superconformal blocks, based on the obtained $c$-recursion relations.
The main idea of \cite{Perlmutter:2015iya}, where an analogues approach has been proposed for Virasoro blocks, was to rearrange
the OPE in a given intermediate channel in order to obtain the sum over modules growing from quasiprimary states (annihilated by $L_1$).
In our case the sum over the given NS module splits into the sum over
 $osp(1|2)$ modules, growing from super-quasiprimary states (annihilated by $L_1$, $G_{\half}$, see Appendix \ref{global}). For 4-point NS blocks it leads to the following decomposition
\be
\label{4ptsol}
F=\sum_{p\in \frac{1}{2}\mathbb{N}_0}z^{p}\chi_p\, G(h\rightarrow h+p)\;,
\ee
where $p$ runs over levels of the quasiprimaries and $G$ is the light block \eqref{4ptglobal}. Note that in this decomposition $\chi_0=1$,
$\chi_{\half}=0$, $\chi_1=0$, because there are no quasiprimary states on the levels $\half$ and $1$, if $h$ is not degenerate. The general coefficients $\chi_p$ can be found from the requirement that the ansatz \eqref{4ptsol} satisfies $c$-recursion constraints \eqref{4pt}. We get the
following expression
\be
\label{chi}
\chi_p=\sum_{j=1}^{[ \frac{p}{2}]}\prod_{\ell=1}^j\sum_{\substack{r_{\ell}\,\ge \,2,\\s_{\ell}\,\ge
\,1}}^{\infty}\gamma_{r_\ell,s_\ell}(c_{eff}^{(\ell)},d_i,h_{eff}^{(\ell)})\;,~\text{with}~\sum_{\ell=1}^j r_\ell
s_\ell=2p~\text{and}~r_i+s_i\in 2\mathbb{N}\;
\ee
and
\be
\gamma_{r,s}(c,d_i,h):=\frac{R_{r,s}(h)}{c-c_{r,s}(h)}\;,
\ee
where $R_{r,s}$ are given in
\eqref{res4pt} and the effective
parameters are:
\be
\begin{aligned}
\label{solvparameters}
h_{eff}^{(\ell)}&:=h+\Delta h^{(\ell-1)}\;,\\
\Delta h^{(\ell)}&:=\sum_{r=1}^l \frac{r_\ell s_\ell}{2}\;,\\
c_{eff}^{(\ell)}&:=c_{r_{\ell-1},s_{\ell-1}}(h_{eff}^{(\ell-1)})\;.\\
\end{aligned}
\ee
These relations allow to compute conformal blocks recursively. We note that the corresponding iteration procedure is more appropriate, as compared to the standard $c$-recursion, for numerical computations of the conformal blocks.\footnote{For the rational values of the central charge $c$, corresponding to minimal models, the numerical recursion requires certain modification, see, e.g. \cite{Javerzat:2018maf}.} In particular, in this version at each level there is no need to keep analytic expressions for the lower levels' coefficients, unlike the original $c$-recursion procedure.

\paragraph{Solution for 1-point blocks on the torus.}
 In the same way as we obtain \eqref{4ptsol} for the spherical 4-point case, we find the solution of the $c$-recursion \eqref{1pttor} on the torus:
\be
\begin{aligned}
\label{1ptsol}
F&=\sum_{p\in \frac{1}{2}\mathbb{N}_0}q^{p}\chi_p\, {}\,G(h\rightarrow h+p)\;,
\end{aligned}
\ee
where $F$ is the torus 1-point block \eqref{torus} and $G$ is the torus light block
\eqref{toruslight}. The coefficients are the following
\be
\chi_p=\sum_{j=1}^{[ \frac{p}{2}]}\prod_{\ell=1}^j\sum_{\substack{r_{\ell}\,\ge \,2,\\s_{\ell}\,\ge
\,1}}^{\infty}\gamma_{r_\ell,s_\ell}(c_{eff}^{(\ell)},d,h_{eff}^{(\ell)})\;,~\text{with}~\sum_{\ell=1}^j r_\ell
s_\ell=2p~\text{and}~r_i+s_i\in 2\mathbb{N}\;,
\ee
\be
\gamma_{r,s}(c,d,h):=\frac{R_{r,s}}{c-c_{r,s}(h)}\;,
\ee
where $R_{r,s}$ are given in \eqref{Rtor} and the effective parameters are defined in
\eqref{solvparameters}. The generalization of this construction to the torus $N$-point block is straightforward.

\section{Discussion}
In this paper we have analyzed the Neveu-Schwarz sector of the $\cN=1$ super-Virasoro CFT. We obtained  $c$-recursion relations for multi-point superconformal blocks on the sphere and on the torus, involving top and down components of primary supermultiplets, which are required for  
constructing multi-point correlation functions in $\cN=1$ CFT minimal models, as well as in the $\cN=1$ supersymmetric Liouville filed theory.
Similarly to the ``standard'' four-point (super-)Virasoro case, the multi-point $c$-recursion is based on  the analysis of the analytic structure, which is characterized by two main ingredients: the singular and the regular parts.

The singular part, which is defined by OPE, is obtained by analyzing superconformal fusion rules. The key point here is that in the multi-point supersymmetric  case the singular part still contains only the contribution of simple poles the $(c-c_{r,s})^{-1}$ and the residues are expressed in a simple manner in terms of the sypersymmetric fusion polynomials. The regular part is governed by the light asymptotic, which can be expressed in terms of global blocks of $osp(1|2)$ algebra and NS vacuum characters. Rather simple representation theory allows to find explicitly the $osp(1|2)$ matrix elements and to reduce the computation of the light blocks (in general topology) to the problem of identification of  the effective plumbing parameterization of the moduli space.

 It is shown that the recursion relations can be effectively rewritten in  terms of the light blocks with shifted values of the
intermediate  conformal dimension parameters, which allows to significantly simplify recursion formulas and makes them more suitable
for numeric computations.


There are several  possible extensions of our results. A natural extension is to
analyze the Ramond sector of $\cN=1$ superconformal theory.
The careful analysis of higher genus cases and, in particular, of the genus-two case is desirable (see, e.g., \cite{Gaberdiel:2010jf,Belin:2017nze,Cho:2017fzo,Keller:2017iql}).
\newpage
\paragraph{Acknowledgements.}
We would like to thank M. Bershtein for useful comments. The research was supported by Foundation for the Advancement of Theoretical
Physics and Mathematics ``Basis''. The work of R.G. has been funded by the  Russian Academic Excellence Project `5-100'. This research
was supported in part by the International Center for Theoretical Sciences (ICTS) during a visit for participating in the program -
Kavli
Asian Winter School (KAWS) on Strings, Particles and Cosmology 2018.

\appendix
\section{General $osp(1|2)$ matrix elements}
\label{global}
Here we work with the global part of NS superalgebra only, that is $osp(1|2)$.
\begin{eqnarray}
\label{NS}
\nonumber
\left[L_n,L_m\right] & = & (n-m)L_{m+n}\,,
\\
\left[L_n,G_k\right] & = &\textstyle{\frac{n-2k}{2}}G_{n+k}\,,
\\
\nonumber
\left\{G_k,G_l\right\} & = & 2L_{k+l}\,,
\end{eqnarray}
where $m,n=-1,0,1$ and $k,l=\textstyle{-\half},\half$. The $osp(1|2)$ highest weight is defined by
\begin{align}
L_{1}\,|d\ra=0\;,
\quad G_{\half}\,|d\ra =0\;.
\end{align}
We use the following identities
\begin{eqnarray}
\nonumber L_1L_{-1}^m|d\rangle&=&(2d+m-1)L_{-1}^{m-1}|d\rangle \,,\\
\nonumber G_{\half}L_{-1}^m|d\rangle&=&mG_{-\half}L_{-1}^{m-1}|d\rangle \,,\\
G_{\half}G_{-\half}L_{-1}^m|d\rangle&=&(2d+m)L_{-1}^m|d\rangle \,,\\
\nonumber L_{1}G_{-\half}L_{-1}^m|d\rangle&=&(2d+2m-1)G_{-\half}L_{-1}^{m-1}|d\rangle \,,
\end{eqnarray}
to derive the full set of global matrix elements. We set the following notation
\be
\rho
\big(G_{-\half}^{\alpha_1}L_{-1}^k\nu_{1},G_{-\half}^{\alpha_2}\nu_2,G_{-\half}^{\alpha_3}L_{-1}^m\nu_{3}\big)=\tau_{k,m}^{\alpha_1,\alpha_2,\alpha_3}(d_1,d_2,d_3)\;,\\
\ee
where the $\tau_{k,m}^{\alpha_1,\alpha_2,\alpha_3}$ are given by
\be
\begin{aligned}\nonumber
\tau_{k,m}^{0,0,0}(d_1,d_2,d_3) &= \sum_{p = 0}^{\min[k,m]} \frac{k!}{p!(k-p)!}(2d_3 +m-1)^{(p)} m^{(p)} \\
&\times(d_2+d_3 - d_1)_{m-p}(d_1 + d_2 -d_3+p-m)_{k-p}\;,\\
\tau_{k,m}^{1,0,0}(d_1,d_2,d_3) &= \sum_{p = 0}^{\min[k,m]} \frac{m!}{p!(m-p)!}(2d_3 +m)^{(p)} k^{(p)} \\
&\times(d_2+d_3 - d_1+\textstyle{\half})_{m-p}(d_1 + d_2 -d_3+\textstyle{\half}-m+p)_{k-p}\;,\\
\end{aligned}
\ee
\be
\begin{aligned}
\tau_{k,m}^{0,1,0}(d_1, d_2,d_3) &= \sum_{p = 0}^{\min[k,m]} \frac{k!}{p!(k-p)!} (2d_3 +m-1)^{(p)} m^{(p)}\\
&\times(d_2+d_3 - d_1+\textstyle{\half})_{m-p}(d_1 + d_2 -d_3+p-m+\textstyle{\half})_{k-p}\;,\\
\tau_{k,m}^{0,0,1}(d_1,d_2, d_3) &= \sum_{p = 0}^{\min[k,m]} \frac{k!}{p!(k-p)!} (2d_3 +m)^{(p)} m^{(p)}\\
&\times(d_2+d_3 - d_1+\textstyle{\half})_{m-p}(d_1 + d_2 -d_3-\textstyle{\half}+p-m)_{k-p}\;,\\
\tau_{k,m}^{1,1,0}( d_1,d_2,d_3) &= \sum_{p = 0}^{\min[k,m]} \frac{k!}{p!(k-p)!} (2d_3 +m-1)^{(p)} m^{(p)}\\
&\times(d_2+d_3 - d_1)_{m-p}(d_1 + d_2 -d_3+p-m+1)_{k-p}(d_1+d_2-d_3)\;,\\
\tau_{k,m}^{1,0,1}( d_1,d_2, d_3) &= \sum_{p = 0}^{\min[k,m]} \frac{k!}{p!(k-p)!} (2d_3 +m)^{(p)} m^{(p)}\\
&\times(d_2+d_3 - d_1)_{m-p}(d_1 + d_2 -d_3+p-m)_{k-p}(d_1-d_2+d_3)\;,\\
\tau_{k,m}^{0,1,1}(d_1, d_2, d_3)& = -\sum_{p = 0}^{\min[k,m]} \frac{k!}{p!(k-p)!} (2d_3 +m)^{(p)} m^{(p)}\\
&\times(d_2+d_3 - d_1)_{m-p+1}(d_1 + d_2 -d_3+p-m)_{k-p}\;,\\
\tau_{k,m}^{1,1,1}(d_1,d_2,d_3) &= \sum_{p = 0}^{\min[k,m]} \frac{k!}{p!(k-p)!} (2d_3 +m)^{(p)} m^{(p)}(d_2+d_3 -
d_1+\textstyle{\half})_{m-p}\\
&\times(d_1 + d_2 -d_3+p-m+\textstyle{\half})_{k-p}(d_1+d_2+d_3+\textstyle{\half})\;,
\end{aligned}
\ee
where $(x)^{(m)}$ and $(x)_{(m)}$ are falling and rising Pochhammer symbols respectively.
Not all of these elements are independent due to the reflection properties \eqref{reflection}. The most general $osp(1|2)$ matrix element
reads
\begin{multline}
\rho \big(G_{-\half}^{\alpha_1}L_{-1}^k\nu_{1},G_{-\half}^{\alpha_2}L_{-1}^m\nu_2,G_{-\half}^{\alpha_3}L_{-1}^n\nu_{3}\big)\\=
(d_1-d_2-d_3+\te{\frac{\alpha_1-\alpha_2-\alpha_3}{2}}+k-m-n)^{(m)}\tau_{k,n}^{\alpha_1,\alpha_2,\alpha_3}(d_1,d_2,d_3)\,.
\end{multline}
\section{Some explicit coefficients}
\label{1tor}
Here we provide the first few coefficients of the torus conformal block with the lower component of the vertex operator  \eqref{torus}:
\be
F_{0}=1\,,\;
F_{1/2}=\!\frac{\rho ( G_{-\half} \nu_h,\nu_d ,G_{-\half} \nu_h)}{\langle G_{-\half}\nu_{h}|G_{-\half}\nu_h\rangle}
=\frac{1}{2h}(2h-d)\,,\; F_{1}=\!\frac{\rho( L_{-1}\nu_h,\nu_d,L_{-1}\nu_h)}{\langle L_{-1}\nu_{h}|L_{-1}\nu_h\rangle}
=\frac{2h+d(d-1)}{2h}
\ee
On the  level $\frac{3}{2}$ we fix the following ordering: $\{G_{-\half}L_{-1},G_{-3/2}\}$. The $\frac{3}{2}$ -level coefficient of the
conformal block is
\be
F_{\frac{3}{2}}=\tr \left(B _{\frac{3}{2}}^{-1}(B _{\frac{3}{2}}+ T_{\frac{3}{2}})\right)\;,
\ee
where
\be
B _{\frac{3}{2}}=\left(
 \begin{array}{cc}
 \dps 2h(2h+1)  & 4h  \\
 4h  & 2h+c\\
\end{array}
\right)\;,
\quad
\dps
T _{\frac{3}{2}}=\left(
 \begin{array}{cc}
 \dps \tau_{1,1}(\tilde h, d, \tilde h)  & -d(d+1) \\
 -d(d+1) & -3d\\
\end{array}
\right)\;.
\quad
\dps
\ee
On the  level 2 two we fix the following ordering: $\{L_{-1}^2,~ L_{-2},~ G_{-\half}G_{-3/2}\}$. The second level coefficient of the
conformal block is
\be
F_{2}=\tr \left(B _{2}^{-1}(B _2+ T_2)\right)\;,
\ee
where
\be
\begin{aligned}
B_2 =\begin{pmatrix}
  4h(2h+1)  & 6h & 8h  \\
 6h  & 4h+\frac{3c}{4} & 3h + \frac{3c}{2}\\
 8h & 3h + \frac{3c}{2} & c(3+2h)+2h(2h-1)\\
\end{pmatrix}\;,\\
T_{2}=
\begin{pmatrix}
d(d-1)(8h+d(d-1)+2) & 2d(d^2-1)& d(d-1)(2+3d)\\
2d(d^2-1) & 4d(d-1) & 6d(d-1)\\
d(d-1)(2+3d) &6d(d-1) & d(11d-8h-9-c)\\
\end{pmatrix}\;.
\end{aligned}
\ee

\section{Plumbing constructions}
\label{Plumbing}
For completeness, we describe here the relation (mostly borrowed from the bosonic case \cite{Cho:2017oxl}) between the local parametrization of the moduli space and the elements of the associated dual diagram. We begin with the plumbing construction associated with Fig.~\bref{Fig:4block}. Accordingly, building blocks are two two-punctured and one-holed spheres. We fix the first sphere to have punctures at 0 and 1 and a hole at $\infty$. The second sphere has punctures at 1 and $\infty$ and a hole at 0. We glue these spheres together by their boundaries via $SL(2,\mathbb{C})$ map. Let us choose the coordinates on the spheres to be $w_1$ and $w_2$, then, the gluing map is $w_2=z w_1$. Thus, we have the sphere with four punctures at $0,1,z,\infty$. The diagram corresponds to the case of $\nu_4,\nu_3,\nu_2,\nu_1$ located at these punctures respectively.

\begin{figure}[H]
\begin{center}
\begin{tikzpicture}
\fill  (0,-0.7) circle (0.3mm);
\fill  (0.3,-0.7) circle (0.3mm);
\fill  (-0.3,-0.7) circle (0.3mm);

\shade[ball color=gray!10] (-3.0,0.0) circle (1.0);
\shade[ball color=gray!10] (-6.0,0.0) circle (1.0);
\shade[ball color=gray!10] (3.0,0.0) circle (1.0);
\shade[ball color=gray!10] (6.0,0.0) circle (1.0);

\fill  (6,0.5) circle (0.3mm);
\fill  (3,0.5) circle (0.3mm);
\fill  (-3,0.5) circle (0.3mm);
\fill  (-6,0.5) circle (0.3mm);

\fill  (6.5,-0.4) circle (0.3mm);
\fill  (-6.5,-0.4) circle (0.3mm);

\draw (5.75,-0.075) node {$ \infty$};
\draw (6.75,-0.075) node {$ 0$};
\draw (2.75,-0.075) node {$ \infty$};
\draw (3.75,-0.075) node {$ 0$};

\draw (-5.3,-0.075) node {$ 0$};
\draw (-3.3,-0.075) node {$ \infty$};
\draw (-2.3,-0.075) node {$ 0$};

\draw (-6.3,-0.075) node {$ \infty$};
\draw (6.2,0.6) node {$ 1$};
\draw (-5.8,0.6) node {$ 1$};
\draw (-2.8,0.6) node {$ 1$};
\draw (3.2,0.6) node {$ 1$};

\draw [line width=0.4pt,dashed] (5.5,-0.4) circle (0.15);
\draw [line width=0.4pt,dashed] (3.5,-0.4) circle (0.15);
\draw [line width=0.4pt,dashed] (2.5,-0.4) circle (0.15);
\draw [line width=0.4pt,dashed] (-5.5,-0.4) circle (0.15);
\draw [line width=0.4pt,dashed] (-3.5,-0.4) circle (0.15);
\draw [line width=0.4pt,dashed] (-2.5,-0.4) circle (0.15);

\draw [line width=0.4pt,dashed] (3.5,-0.25) -- (5.5,-0.25);
\draw [line width=0.4pt,dashed] (3.5,-0.55) -- (5.5,-0.55);
\draw [line width=0.4pt,dashed] (-3.5,-0.25) -- (-5.5,-0.25);
\draw [line width=0.4pt,dashed] (-3.5,-0.55) -- (-5.5,-0.55);

\draw [line width=0.4pt,dashed] (2.5,-0.25) -- (1.5,-0.25);
\draw [line width=0.4pt,dashed] (2.5,-0.55) -- (1.5,-0.55);
\draw [line width=0.4pt,dashed] (-2.5,-0.25) -- (-1.5,-0.25);
\draw [line width=0.4pt,dashed] (-2.5,-0.55) -- (-1.5,-0.55);
\end{tikzpicture}
\end{center}
\caption{The plumbing construction for the $N$-punctured sphere.}
\label{Fig:Nplumb}
\end{figure}
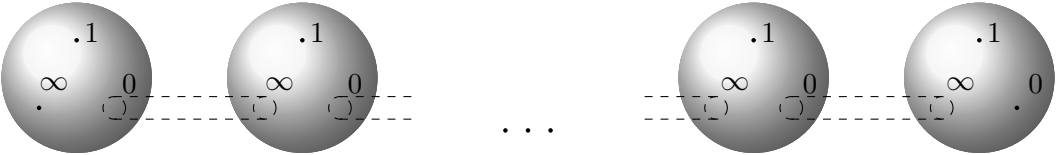

We proceed with the case represented in Fig. \bref{Fig:Nblock}. We have $N-2$ spheres: two two-punctured and one-holed spheres and $N-4$ two-holed and one-punctured spheres. The gluing of the holes is depicted in Fig.~\bref{Fig:Nplumb}.
\noindent We introduce coordinates $w_1$,..., $w_{N-2}$ on the spheres  and the gluing maps $w_i=q_iw_{i+1}$, where $1\le i \le N-3$. We get the sphere with $N$ punctures at $0,\,1,\,q_1,\,q_1q_2,\,...,\, q_1...q_{N-3},\,\infty$. Inserting $\nu_N,\nu_{N-1},...,\nu_1$ at these punctures respectively we come to \eqref{Nmodules}.

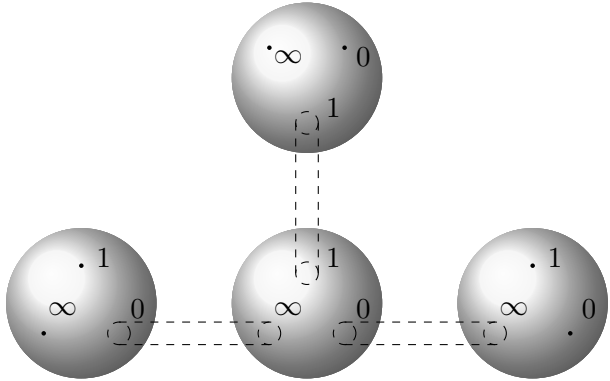
\begin{figure}[H]
\begin{center}
\begin{tikzpicture}[rounded corners,thin]
\shade[ball color=gray!10] (0.0,0.0) circle (1.0);
\shade[ball color=gray!10] (3.0,0.0) circle (1.0);
\shade[ball color=gray!10] (-3.0,0.0) circle (1.0);
\shade[ball color=gray!10] (0.0,3.0) circle (1.0);
\draw [line width=0.4pt,dashed] (0.5,-0.4) circle (0.15);
\draw [line width=0.4pt,dashed] (-0.5,-0.4) circle (0.15);
\draw [line width=0.4pt,dashed] (2.5,-0.4) circle (0.15);
\draw [line width=0.4pt,dashed] (-2.5,-0.4) circle (0.15);
\draw [line width=0.4pt,dashed] (0,0.4) circle (0.15);
\draw [line width=0.4pt,dashed] (0,2.4) circle (0.15);

\draw [line width=0.4pt,dashed] (0.15,0.4) -- (0.15,2.4);
\draw [line width=0.4pt,dashed] (-0.15,0.4) -- (-0.15,2.4);
\draw [line width=0.4pt,dashed] (0.5,-0.25) -- (2.5,-0.25);
\draw [line width=0.4pt,dashed] (0.5,-0.55) -- (2.5,-0.55);
\draw [line width=0.4pt,dashed] (-0.5,-0.25) -- (-2.5,-0.25);
\draw [line width=0.4pt,dashed] (-0.5,-0.55) -- (-2.5,-0.55);

\fill  (3,0.5) circle (0.3mm);
\fill  (-3,0.5) circle (0.3mm);
\fill  (3.5,-0.4) circle (0.3mm);
\fill  (-3.5,-0.4) circle (0.3mm);
\fill  (0.5,3.4) circle (0.3mm);
\fill  (-0.5,3.4) circle (0.3mm);

\draw (3.75,-0.075) node {$ 0$};
\draw (2.75,-0.075) node {$ \infty$};
\draw (0.75,-0.075) node {$ 0$};
\draw (-0.25,-0.075) node {$ \infty$};
\draw (-2.25,-0.075) node {$ 0$};
\draw (-3.25,-0.075) node {$ \infty$};
\draw (0.35,0.6) node {$ 1$};
\draw (0.35,2.6) node {$ 1$};
\draw (3.3,0.6) node {$ 1$};
\draw (-2.7,0.6) node {$ 1$};
\draw (0.75,3.275) node {$ 0$};
\draw (-0.25,3.275) node {$ \infty$};
\end{tikzpicture}
\end{center}
\caption{The plumbing construction for the 6-point non-linear block.}
\label{Fig:plumb6pt}
\end{figure}

Now we consider 6-point block with the dual diagram in Fig.~\bref{Fig:6block}. The corresponding plumbing graph is shown in Fig. \bref{Fig:plumb6pt}. It consists of three two-punctured and one-holed spheres and one three-holed sphere.
\noindent Introducing coordinates $w_i$, $i=1,2,3$ on two-punctured spheres numerated from left to right and $w_4$ on the three-punctured sphere in the middle one, we get the following maps
\be
w_1=q_2 w_4\,,\quad w_2=1+\frac{q_1}{w_4-1}\,,\quad w_3=\frac{w_4}{q_3}\;.
\ee
Gluing via these mappings we get the sphere with 6 punctures located at points: \\$0,\,1-q_1,\,\frac{1}{q_2},\,q_3,\,1,\,\infty$.

\begin{figure}[H]
\begin{center}
\begin{tikzpicture}[rounded corners,thin]
\shade[ball color=gray!10] (0.0,0.0) circle (1.0);
\fill  (0,0.5) circle (0.3mm);
\draw [line width=0.4pt,dashed] (0.5,-0.4) circle (0.15);
\draw [line width=0.4pt,dashed] (-0.5,-0.4) circle (0.15);
\draw[black,dashed] (-0.65,-0.4) to [out=270,in=180] (0.0,-1.5);
\draw[black,dashed] (0.0,-1.5) to [out=0,in=-90] (0.65,-0.4);
\draw[black,dashed] (-0.35,-0.4) to [out=270,in=180] (0.0,-1.25);
\draw[black,dashed] (0.0,-1.25) to [out=0,in=-90] (0.35,-0.4);
\draw (0.65,-0.075) node {$ 0$};
\draw (0.15,0.675) node {$ 1$};
\draw (-0.45,-0.075) node {$ \infty$};
\end{tikzpicture}
\end{center}
\caption{The plumbing construction for the 1-point torus block.}
\label{Fig:plumb1pttor}
\end{figure}
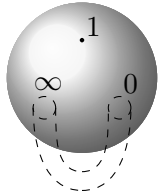

For the torus one-point block (Fig.~\bref{Fig:1torusblock}) we have the plumbing construction in Fig.~\bref{Fig:plumb1pttor}.
\noindent One-punctured torus is obtained from two-holed one-punctured sphere by identification \\$w\sim qw$ and $q$ is the modulus of the given torus.

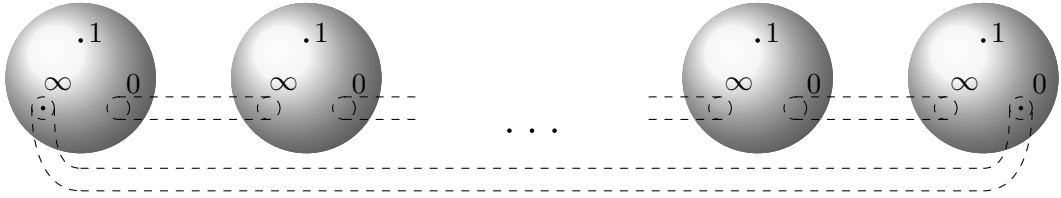
\begin{figure}[H]
\begin{center}
\begin{tikzpicture}
\fill  (0,-0.7) circle (0.3mm);
\fill  (0.3,-0.7) circle (0.3mm);
\fill  (-0.3,-0.7) circle (0.3mm);

\shade[ball color=gray!10] (-3.0,0.0) circle (1.0);
\shade[ball color=gray!10] (-6.0,0.0) circle (1.0);
\shade[ball color=gray!10] (3.0,0.0) circle (1.0);
\shade[ball color=gray!10] (6.0,0.0) circle (1.0);

\fill  (6,0.5) circle (0.3mm);
\fill  (3,0.5) circle (0.3mm);
\fill  (-3,0.5) circle (0.3mm);
\fill  (-6,0.5) circle (0.3mm);

\fill  (6.5,-0.4) circle (0.3mm);
\fill  (-6.5,-0.4) circle (0.3mm);

\draw (5.75,-0.075) node {$ \infty$};
\draw (6.75,-0.075) node {$ 0$};
\draw (2.75,-0.075) node {$ \infty$};
\draw (3.75,-0.075) node {$ 0$};

\draw (-5.3,-0.075) node {$ 0$};
\draw (-3.3,-0.075) node {$ \infty$};
\draw (-2.3,-0.075) node {$ 0$};

\draw (-6.3,-0.075) node {$ \infty$};
\draw (6.2,0.6) node {$ 1$};
\draw (-5.8,0.6) node {$ 1$};
\draw (-2.8,0.6) node {$ 1$};
\draw (3.2,0.6) node {$ 1$};

\draw [line width=0.4pt,dashed] (5.5,-0.4) circle (0.15);
\draw [line width=0.4pt,dashed] (3.5,-0.4) circle (0.15);
\draw [line width=0.4pt,dashed] (2.5,-0.4) circle (0.15);
\draw [line width=0.4pt,dashed] (-5.5,-0.4) circle (0.15);
\draw [line width=0.4pt,dashed] (-3.5,-0.4) circle (0.15);
\draw [line width=0.4pt,dashed] (-2.5,-0.4) circle (0.15);
\draw [line width=0.4pt,dashed] (-6.5,-0.4) circle (0.15);
\draw [line width=0.4pt,dashed] (6.5,-0.4) circle (0.15);

\draw [line width=0.4pt,dashed] (3.5,-0.25) -- (5.5,-0.25);
\draw [line width=0.4pt,dashed] (3.5,-0.55) -- (5.5,-0.55);
\draw [line width=0.4pt,dashed] (-3.5,-0.25) -- (-5.5,-0.25);
\draw [line width=0.4pt,dashed] (-3.5,-0.55) -- (-5.5,-0.55);

\draw [line width=0.4pt,dashed] (2.5,-0.25) -- (1.5,-0.25);
\draw [line width=0.4pt,dashed] (2.5,-0.55) -- (1.5,-0.55);
\draw [line width=0.4pt,dashed] (-2.5,-0.25) -- (-1.5,-0.25);
\draw [line width=0.4pt,dashed] (-2.5,-0.55) -- (-1.5,-0.55);

\draw[black,dashed] (-6.65,-0.4) to [out=270,in=180] (-6.0,-1.5);
\draw[black,dashed] (-6.35,-0.4) to [out=270,in=180] (-6.0,-1.20);
\draw[black,dashed] (6.0,-1.5) to [out=0,in=-90] (6.65,-0.4);
\draw[black,dashed] (6.0,-1.2) to [out=0,in=-90] (6.35,-0.4);

\draw [line width=0.4pt,dashed] (-6,-1.5) -- (6,-1.5);
\draw [line width=0.4pt,dashed] (-6,-1.20) -- (6,-1.20);
\end{tikzpicture}
\end{center}
\caption{The plumbing construction for the $N$-punctured torus.}
\label{Fig:Nplumbtorus}
\end{figure}

The dual diagram for the torus $N$-point block in the necklace channel (Fig.~\bref{Fig:Ntorusblock}) corresponds to the plumbing construction depicted in Fig.~\bref{Fig:Nplumbtorus}.
\noindent We obtain the torus with modulus $q=q_1...q_{N}$ and  punctures located at $1,\,q_1,\,q_1q_2,\,...,\, q_1...q_{N-1}$.

\bibliographystyle{JHEP}
\providecommand{\href}[2]{#2}\begingroup\raggedright\endgroup

\end{document}